\documentstyle[bibnorm]{lamuphys2}
\begin{document}
\title{On Heterotic/Type I Duality in $d=8$ $^*$}

\author{Kristin\,F\"orger}

\institute{Centre de Physique Th\'{e}orique, Ecole Polytechnique,
F-91128 Palaiseau, France}

\maketitle
\renewcommand{\thefootnote}{\fnsymbol{footnote}}
\footnotetext[1]{To appear 
in the proceedings of {\it Quantum Aspects of Gauge
Theories, Supersymmetry and Unification}, Corfu, September 1998. 
\hfill CPHT-PC696.1298}
\renewcommand{\thefootnote}{\arabic{footnote}}
\begin{abstract}
We discuss heterotic corrections to quartic 
internal $U(1)$ gauge couplings and check duality by calculating 
one-loop open string diagrams and identifying the $D$-instanton sum
in the dual type I picture. We also compute $SO(8)^4$ 
threshold corrections and finally $R^2$ corrections in type I theory. 
\end{abstract}
%
\section{Introduction}
Heterotic $SO(32)$/type \ I duality relates 
two string theories in ten dimensions \cite{wie,pw}. 
A field redefinition transforms the low energy effective action of the
heterotic string into the one of the type I string \cite{pw,ts}
\begin{equation}\label{dualmap}
G_{\mu\nu}^I= \lambda^I G_{\mu\nu}^{\rm het}\, ,\quad
\lambda^I=\frac{1}{\lambda^{\rm het}}\,,\quad
B_{\mu\nu}^{R,I}=B_{\mu\nu}^{NS,{\rm het}}
\end{equation}
where $G_{\mu\nu}$ is the metric,
$B_{\mu\nu}^{NS/R}$ the antisymmetric tensor in the $NS$ or $R$ sector
and $\lambda^I=e^{\phi^{(10)}}$ the ten dimensional type I coupling.
Compactifying to lower dimensions
modifies the duality map due to the dependence on the
volume of the compactification manifold.

Besides the equivalence of the BPS spectrum of the heterotic and
type I string, non--trivial checks can be done for the BPS--saturated 
${\cal F}^4$ and ${\cal R}^4$ couplings; see e.g. \cite{bk,BFKOV,ko,BGMN}  
and \cite{LS} for gauge couplings in the context of  
heterotic/F-theory duality.
The special feature of these couplings is that --on the heterotic side-- 
they are protected from higher than one--loop corrections in the
effective action thanks to supersymmetry. Thus they can be exactly
calculated.
These threshold corrections translate to perturbative open string amplitudes
and non-perturbative BPS $D1$-instanton corrections on  the type\ I side.
Therefore by studying various examples of couplings 
one learns more about the rules of computing 
non-perturbative $D$-brane contributions (see e.g. \cite{KV}).

Toroidal compactification down to eight dimensions
gives rise to $4$ abelian gauge fields, 
corresponding to the components $G_{\mu I}$ and $B_{\mu
I}$, where $\mu$ is the space-time index and $I$ labels the
compact directions.
The four real massless scalar fields parametrize the moduli space
$\frac{O(2,2)}{O(2)\times O(2)}$ of the torus $T^2$ 
(if no Wilson lines are switched on).

In the type I theory the $U(1)^2\times U(1)^2$ gauge group is
reduced to a diagonal $U(1)^2$, due to the twist operator 
$\Omega$ \cite{bisa}. Similarly only two of the four massless scalar 
fields survive the projection.
\section{Heterotic string in eight dimensions}
At string one--loop, the higher derivative couplings
${\cal F}^4, ({\cal F}^2)^2, {\cal R}^4$ and $({\cal R}^2)^2$ and 
their respective CP--odd parts receive corrections
in the effective string action. These constant contributions are highly fixed
by ten--dimensional anomaly cancellation arguments.
In the Green-Schwarz formalism they are calculated by (almost) holomorphic 
one loop string amplitudes, whose 
minimal number of external legs is fixed by 
saturating fermionic zero modes.
The result for the amplitudes  is summarized by
the worldsheet $\tau$ integral over a weight zero
almost holomorphic function which is related 
to the elliptic genus \cite{lsw,WL}.
After compactification on $T^2$ these corrections
become moduli dependent functions. It is believed that there are no
space--time instanton effects, since the only supersymmetric soliton,
the NS five--brane, cannot be wrapped around the torus.  
\subsection{$U(1)_{\rm int}^2$ heterotic threshold corrections} 
Here we are mainly interested in threshold corrections to internal
$U(1)$ couplings. At tree level  
they are obtained by dimensional reduction of the ${\cal R}^2$ term
in the effective action.

The internal gauge boson vertex operator
can be read off from the $\sigma$ model action.
The internal metric is
$G_{IJ}=\epsilon_{(I}^{+} \epsilon_{J)}^{-}$ with $\epsilon_I^+=
\sqrt\frac{T_2}{U_2}(1,U)$ 
and $\epsilon_J^-=\sqrt\frac{T_2}{U_2} (1,\bar U)$. Therefore 
the vertex operator for $G_{\mu I}$ gauge bosons 
in the background gauge 
$\epsilon_{\mu}e^{i k\cdot X}=-\frac{1}{2} F_{\mu\rho} X^{\rho}$ with 
$F_{\mu\nu}=\rm const$ in Green-Schwarz formalism is:
\begin{equation}\label{vgauge}
V_{\rm gauge}=\frac{i\pi}{\tau_2} F_{\mu\nu}^A\, Q^A
:\Big(\partial X^\mu X^\nu-\frac{1}{4}
\gamma_{ab}^{\mu\nu} S^a S^b\Big): \, ,
\end{equation}
where $A=\pm$ labels different $U(1)$ charges
$Q^+=\epsilon^+_I\bar\partial X^I=\sqrt\frac{T_2}{U_2} (1,U) A 
({\tau\atop 1})$ and $Q^-$, which is
obtained from $Q^+$ by replacing $U$ by $\bar U$
and $A=\Big({n_1\atop n_2} {-l_1\atop -l_2}\Big)$ 
is a $GL(2,{\bf Z})$ matrix. 
The complex structure and the K\"ahler
modulus of the torus are defined in terms of the metric
and NS-NS antisymmetric tensor as $U=U_1+i U_2=(G_{12}+i\sqrt{\det 
G})/G_{11}$
and $T=T_1+iT_2=2 (b+i\sqrt{\det G})$, respectively.

The effective action which arises from string amplitudes 
with vertex operators (\ref{vgauge}) is:
\begin{equation}\label{hamp}
{\cal I}^{\rm het}_{\rm 1-loop} = \frac{V^{(8)} T_2}{2^8\pi^4}
\int \frac{d^2\tau}{\tau_2^2} 
\sum_{{\bf l},{\bf n}\in {\bf Z}}e^{-\frac{2\pi}{\tau_2}
(n^I\tau-l^I)(G+B)_{IJ}(n^J{\bar\tau}- l^J)}
\int DS^a_0 \ e^{{-\pi/\tau_2}
F_{\mu\nu}^A\ Q^A R_0^{\mu\nu}} \bar{\cal A}(\bar q)\, , 
\end{equation}
where $R_0^{\mu\nu}=\frac{1}{4}S_0^a \gamma_{ab}^{\mu\nu} S_0^b$ and
${\cal A}(\bar q)=\frac{{\bar E}_4^2}{\bar\eta^{24}}$.\footnote{
Supersymmetry relates even to  odd spin structures. Since periodic
Green-Schwarz fields $S^a$ are mapped to periodic NSR fields $\psi^\mu$, 
CP-odd correlation functions of NSR currents are equivalent to
CP-even correlation functions of Green-Schwarz currents due to
a Riemann identity \cite{WL}.}  
Berezin integration thus produces the following terms in the
effective action
\begin{equation}\label{ihet}
{{\cal I}}^{\rm het}_{\rm 1-loop}=\frac{V^{(8)}T_2}{2^8 \pi^4}
\Delta_{F_+^{4-l} F_-^l}t_8 F_+^{4-l} F_-^l\, ,
\end{equation}
for $l=0,\ldots,4$ with coupling
\begin{equation}\label{deltaa}
\Delta_{F_+^{4-l} F_-^l}= 
\frac{\partial^{4-l}}{\partial {s_+}^{4-l}}\frac{\partial^{l}}
{\partial {s_-}^{l}}\int \frac{d^2\tau}{\tau_2^{2}} \sum_{A\in GL(2,{\bf Z})}
e^{2\pi i T \det A-\frac{\pi T_2}{\tau_2 U_2}|(1,U) 
A\left(\tau\atop -1\right)|^2}
e^{- s_A \frac{\pi}{\tau_2} Q^A}\bar{\cal A}(\bar q)
\Big|_{s=0}\, .
\end{equation} 
Poisson resummation turns the sum over winding modes ${\bf l}, {\bf
n}$ into
a sum over momenta ${\bf m}$ and windings ${\bf n}$, thereby transforming
the charge insertions $Q^\pm$ into Narain momenta insertions 
$P_{R/L}$.

The $\tau$-integrals can be performed \cite{LS,BGMN,FS2} by the unfolding
trick \cite{DKLII}  and techniques developed in \cite{fs}.
Unfolding the integration region into
the trivial orbit $A=0$
the degenerate orbit $A=\left( {0\atop 0}{j\atop p}\right)$ with $(j,p)\neq 0$
and the non-degenerate orbit $A=\left( {k \atop 0}{j\atop p}\right)$ with
$0\le j<k$ and $p>0$ enables one to identify these contributions
to tree-level, perturbative and
non-perturbative corrections on the type I side.
The full heterotic effective action can be expressed in terms of 
prepotentials \cite{ko,LS}. 
This is similar to the case of $N=2$ supersymmetric string vacua in $d=4$. 

Integration of the non-degenrate orbit of (\ref{deltaa}) gives \cite{FS2}
\begin{equation}\label{intFV}
\Delta_{F_+^{4-l} F_-^l}^{\rm non-deg}=
\frac{\partial^{4-l}}{\partial {s_+}^{4-l}}\frac{\partial^{l}}
{\partial {s_-}^{l}}
\sum_{{0\le j<k\atop p\neq 0}}\frac{1}{k\sqrt{b(s_+,s_-)}} 
e^{-2\pi i  k p {\tilde T}(s_+,s_-)}
{\cal A}\Big({\tilde U}(s_+,s_-)\Big)\, ,
\end{equation}
where 
\begin{eqnarray}\label{btu}
b(s_+,s_-)&=&\Big(p-\frac{i(s_++s_-)}{2\pi\sqrt{2T_2 U_2}}\Big)^2+
\frac{s_+ s_-}{2\pi^2 T_2 U_2}\nonumber\\
\tilde T(s_+,s_-)&=& T_1-\frac{i}{p} T_2
\Big[\sqrt{b(s_+,s_-)}+\frac{i(s_+-s_-)}{2\pi\sqrt{2 T_2 U_2}}\Big]\\ 
\tilde U(s_+,s_-)&=&\frac{1}{k}
\Big(j+p U_1-i U_2\Big[\sqrt{b(s_+,s_-)}+\frac{i(s_+-s_-)}{2\pi\sqrt{2
T_2 U_2}}\Big]\Big)\ .\nonumber
\end{eqnarray}
The worldsheet instanton corrections
$\Delta^{\rm non-deg}$ are exponentially suppressed.
They can be expressed in terms of Hecke operators:\footnote{The 
Hecke operator acts on a modular form
$\Phi_w$ of weight $w$ as $H_N[\Phi_w](U)=\frac{1}
{N^{1-w}}\sum_{k,p>0\atop kp=N}\sum_{0\le j<k} k^{-w}\Phi_w({\cal U})$ with
the complex structure ${\cal U}=\frac{j+p U}{k}$ \cite{Serre}.}
\begin{eqnarray}\label{instFV}
\Delta_{F_+^{4-l}F_-^l}^{\rm non-deg}&=&\frac{\pi^4}{2}
\sum_{N=1}^\infty\bigg[\Big(\frac{T_2}{U_2}\Big)^{2-l}
\Big(D_{{\cal T}}^{4-l}q_{{\cal T}}\Big) 
N^{4-l} H_N[D_U^l{\cal A}](U)\\
&+&\Big(\frac{U_2}{T_2}\Big)^{2-l}
D_{\bar {\cal T}}^l\bar q_{\bar {\cal T}} \, \frac{1}{N^{4-l}} \, 
H_N[\bar D_U^{4-l}\bar {\cal A}](\bar U)\bigg]\; ,\nonumber
\end{eqnarray}
where $q_{{\cal T}}=e^{2\pi i{\cal T}}$ with 
${\cal T}=N T$.\footnote{The covariant derivative 
$D_w$ which maps modular forms of weight 
$(w,\bar w)$ to forms of
weight $(w+2,\bar w)$ is defined
by $\phi_{w+2,\bar w}(U)=D_w\phi_{w,\bar w}(U)
 =\frac{i}{\pi}\Big(\partial_U+\frac{w}{(U-\bar U) }\Big)\phi_{w,\bar w}(U)$.
We use the notation $D^k\Phi_{w,\bar w}=D_{w+2 (k-1)}
D_{w+2(k-2)}\ldots D_w \Phi_{w,\bar w}$.}
This rewriting is one step towards the understanding
of semi-classical $D$-instanton calculus in the dual type I picture.

For the degenerate orbit we obtain  
\begin{equation}\label{degTTTT}
\Delta^{\rm deg}_{F_+^{4-l} F_-^l}=\frac{c_0}{\pi}\Gamma(5)
\frac{U_2^3}{T_2^3}
\sum_{(j,p)\neq (0,0)}\frac{(j-p\, U)^{4-l}(j-p\, \bar U)^{l}}{|j-p
  U|^{10}}\ ,
\end{equation}
where $c_0=504$ originates from $\frac{{\bar E}_4^2}{{\bar
\eta}^{24}}=\sum_n c_n \bar q^n$. 
The sum is taken over winding modes $j,p$. 
Setting $n_1=n_2=0$ corresponds to 
vanishing winding numbers. 
Only Kaluza-Klein momenta contribute to $\Delta^{\rm deg}$, 
which will be identified with type I perturbative corrections in the
next section. For $l=2$ the result can be expressed in terms of
the generalized Eisenstein function 
$E(U,3)=\sum_{(j.p)\neq 0}\frac{U_2^3\zeta(6)}{|j-p U|^6}$ \cite{Terras}.
\section{Type\ I in eight dimensions}

In the following we calculate two kinds of one--loop gauge
threshold corrections:
First corrections w.r.t. the internal gauge group $U(1)_{\rm int}^2$ 
and second corrections including discrete Wilson lines that break the gauge
group to $SO(8)^4$. Finally we discuss type I $R^2$ corrections.

\subsection{$U(1)_{\rm int}^2$ type I threshold corrections}
Using the duality map (\ref{dualmap}) 
one can verify that the degenerate orbit of the heterotic threshold 
corrections corresponds to one loop terms in the type I effective action: 
\begin{equation}\label{hetdeg}
{\cal I}^{\rm het}_{\rm deg}=\frac{V^{(8)} T_2}{2^8\pi^4} 
t_8 F_+^{4-l}F_-^l\Delta^{\rm deg}_{F_+^{4-l} F_-^l}(T_2,U)
\leftrightarrow{\cal  I}^{\rm I}_{\rm 1-loop}\, .
\end{equation}
$V^{(8)} t_8$ is invariant under this transformation,
and the factor $e^{2\phi^I}$ which arises  from
$\frac{T_2}{T_2^3}$ is canceled by 
$G_{\rm int}^{-2}$ which contracts internal indices of the gauge
kinetic term, altogether resulting in a $\lambda_I^0$ coupling.
In the following we will check this by an independent calculation
of the type I one-loop corrections to these couplings.

One loop open string amplitudes consist in
summing over oriented and unoriented surfaces 
with and without boundaries like the
torus (${\cal T}$) and Klein bottle (${\cal K}$) 
for the closed string sector and
the annulus (${\cal A}$) and M{\"o}bius  strip 
(${\cal M}$) for the open string sector,
which have Euler number $\chi=0$.

The vertex operator of the gauge fields coincides with
the one of the type IIB theory:
\begin{equation}\label{vert}
V_{\rm gauge}=G_{\mu I} :(\bar \partial X^I-\frac{1}{4} 
k_\sigma \bar S^a \gamma^{I\sigma}_{a,b}\bar S^b)
(\partial X^\mu-\frac{1}{4}k_\nu S^a\gamma^{\nu\mu}_{ab}S^b) e^{i k
  X}:\, .
\end{equation}

There is no contribution from the torus diagram ${\cal T}$ since the sixteen
fermionic zero modes cannot be saturated at the level of four
derivative terms. 
The remaining amplitudes can be written as:
\begin{equation}\label{oampl}
{\cal I}_{\rm 1-loop}^I=\frac{1}{2} V^{(8)}\sum_{\sigma}\rho_{\sigma}
\int_0^\infty \frac{dt}{t}\frac{1}{(2\pi^2 t)^4}
\Big(\sum_{p\in \Gamma_2}e^{-\pi t |p|^2/2}\Big)\  Z(\tau_{\sigma})
\int\prod_{i=1}^4 d^2 z_i\langle\prod_{i=1}^4 {V}_{{\rm gauge},i}\rangle
\end{equation}
where the sum is taken over one-loop surfaces 
$\sigma={\cal A},{\cal M},{\cal K}$
with relative weights $\rho_{{\cal K}}=1$, $\rho_{{\cal A}}=N^2$ and
$\rho_{{\cal M}}=-N$, and $N$ is the Chan-Paton charge which takes the
value $N=32$ for the gauge group  $SO(32)$.
The factor $(2\pi^2 t)^{-4}$ arises from momentum integration and
$V^{(8)}$ is the uncompactified volume in type I units.
The open string oscillator sum is
$Z(\tau_{\sigma})=\frac{1}{\eta^{12}(\tau_{\sigma})}\sum_{\alpha=2,3,4}\frac{1}{2}
s_\alpha\theta_\alpha^4(0,\tau_{\sigma})$
with GSO projection signs 
$s_3=-s_2=-s_4=1$ and modular parameters
$\tau_{{\cal A}}=\frac{it}{ 2}, \tau_{{\cal M}}=\frac{it+1}{ 2}, 
\tau_{{\cal K}}={2it}$.
 The torus partition function $\Gamma_2$ is now
restricted to the Kaluza-Klein momenta $m_1$ and $m_2$ and
$p^2=p_I G^{IJ} p_J=\frac{1}{2 U_2 \sqrt{G}}|m_1+m_2 U|^2$. This reflects
the fact that
for the open string the perturbative duality group is reduced to
$SL(2,{\bf Z})_U$.

Contraction of the leftmoving fermions $S^a\gamma^{\nu\mu}_{ab} S^b$
contributes  four derivatives to the amplitude and  
using a Riemann identity one finds
$Z(t)G_F^4(t)=-\frac{\pi^4}{2}$
which is independent of worldsheet coordinates.
Poisson resummation gives:
\begin{equation}\label{typeIthr}
{\cal I}^I_{\rm 1-loop}=\frac{V^{(8)} T_2}{2^8}  t_8 F_+^{4-l} F_-^l
\int \frac{dt}{t^6}\sum_{j,p\neq 0}
e^{\frac{-\pi T_2}{t U_2}|j-p U|^2} \frac{T_2^2}{U_2^2} (j-p U)^{4-l} 
(j-p \bar U)^l \frac{1}{2}\Big(N^2-N+2^4\Big)
\end{equation}
where $\frac{1}{2} \Big(N^2-N+2^4\Big)=c_0$.
Integration over $t$ thus reproduces the corresponding heterotic coupling
$\Delta^{\rm deg}_{F_+^{4-l} F_-^l}(T_2,U)$.
\subsection{$SO(8)^4$ type I threshold corrections}
This is the orientifold example of Sen \cite{sen}, for which
heterotic-F theory duality can be checked explicitly \cite{LS}.
We switch on discrete Wilson lines $a_1^I=\frac{1}{2}(0^4,0^4,1^4,1^4)$ and 
$a_2^I=\frac{1}{2} (0^4,1^4,0^4,1^4)$
which break the $SO(32)$ gauge group to $SO(8)^4$. 
The internal $U(1)^2$ gauge group cannot be 
enhanced for this particular choice of Wilson lines \cite{pw}
and the underlying prepotenial is trivial. 
I.e. in this case the corresponding
gauge couplings $\Delta_{F_+^{4-l}F_-^l}$ vanish identically \cite{LS}.

We apply the background field method of
\cite{BP,bk} to calculate type I one-loop threshold corrections.
The expression for the one-loop amplitude reads:
\begin{eqnarray}\label{amplbg}
{\cal I}_{\rm 1-loop}^I & =&\frac{i V^{(8)}}{ 2} 
\sum_{\sigma, ij}\rho_{\sigma}  \int \frac{dt}{t} \frac{1}{(2 \pi^2 t)^4}
\sum_{a_{ij}^{\sigma}+\Gamma_2} e^{-\pi t p^2/2} \frac{1}{\eta^{12}({
\tau_\sigma})}
 \frac{i}{2} q_{ij}^\sigma B t 
\frac{\theta_1'(0,\tau_{\sigma})}{\theta_1\Big(\frac{i\epsilon_{\sigma} t}{2},
\tau_\sigma\Big)}\\
 && \sum_{\alpha}\frac{1}{2} s_{\alpha}\theta_{\alpha}
\Big(\frac{i\epsilon_{\sigma} t}
{2},\tau_\sigma\Big)
\theta^3_{\alpha}(0,\tau_{\sigma})\nonumber
\end{eqnarray}
where 
$F=B Q$ is the background gauge field and $Q$ a generator of
the Cartan subalgebra and $q_i$ the corresponding charge.
The non-linear function $\epsilon_\sigma$ can be expanded as 
$\epsilon_\sigma\simeq q_{\sigma} B+{\cal O}(B^3)$ with 
$q_{ij}^{{\cal A}}=(q_i+q_j)$ and 
$q^{{\cal M}}_{ij}=2 q_i$. 

Expanding the integrand to the order ${\cal O}(B^4)$ gives
\begin{equation}\label{anmoeb}
{\cal I}_{\rm 1-loop}^I=-\frac{V^{(8)}}{2} \sum_{\sigma, ij} 
\rho_\sigma  \int
\frac{dt}{t}
\Big[\epsilon_{\sigma}^4
\sum_{a_{ij}^\sigma+\Gamma_2}e^{-\pi t p_I G^{IJ} p_J/2}\Big]
\end{equation}
After Poisson resummation and changing variables from the
direct channel to the closed string transverse channel
$l=1/t$ for the annulus and $l=1/(4t)$ for the M\"obius strip,
one finds:
\begin{equation}\label{acmc}
{\cal I}^I_{\rm 1-loop}=-\frac{iV^{(8)} B^4}{2\pi} 
\sum_w\frac{T_2}{w^I G_{IJ} w^J}\bigg[ \sum_{ij}
e^{2\pi i (a_i+a_j)_I w^I}(q_i+q_j)^4
-\sum_i e^{4\pi i a_{iI} w^I}(2 q_i)^4\bigg]
\end{equation}
Evaluating the sum leads to
\begin{eqnarray}\label{sectors}
{\cal I}^I&=&i\pi V^{(8)}B^4\bigg[4\ln\Big[T_2 U_2|\eta(U)|^4\Big]
\sum_{i<j; (i,j)=({\bf k},{\bf k})}(q_i+q_j)^4\nonumber\\
&&+ 2 \sum_{k={\bf 2}}^{\bf 4}
\ln\Big[T_2 U_2\Big|\frac{ \theta_k(U)}{2 \eta(U)}\Big|^2\Big]
\sum_{(i,j)=({\bf 1},{\bf k})}(q_i+q_j)^4\bigg] 
\end{eqnarray}
where ${\bf 1}=\{1,\ldots,4\}, {\bf 2}=\{5,\ldots,8\},{\bf
4}=\{9,\ldots,12\}, {\bf 3}=\{13,\ldots,16\}$.
We omitted some moduli 
independent constant which appears after regularization of the
logarithmic divergence \cite{DKLII}.

In the T-dual type I' picture this example corresponds to
placing four seven branes at each of the
four fixed points. The above result are the threshold
corrections to ${\rm Tr}F^2_{SO(8)_k} {\rm Tr} F^2_{SO(8)_{k'}}$
for $k,k'=1,\ldots,4$ coming from open strings stretched 
between the branes sitting on the same fixed point or different fixed points.
On the heterotic side they translate to the degenerate orbit of the
coupling \cite{LS}. 
\subsection{D--instanton contribution}
Heterotic worldsheet instantons that appear 
in the non-degenerate orbit of the heterotic amplitude,
are 'dual' to type I BPS $D$--instantons which 
are wrapped $D1$ branes on the spacetime torus. 
The way to count all inequivalent ways in which a torus 
can cover $N$-times  the space-time torus is characterized
by the transformations matrix $A$ of the non-degenerate orbit
where $kp=N$ is the instanton number \cite{BFKOV}.   

The classical instanton saddle point
is the exponent of the  Born-Infeld action 
of the wrapped $D1$-brane \cite{BFKOV,ko,BGMN} 
\begin{equation}
S_{BI}=\int d^2\sigma \frac{1}{\lambda_I} \sqrt{\det{\hat G}_I}-i\int
\hat B_I\, .
\end{equation}
In the presence of $U(1)$ background fields the square root also  includes
gauge fields $\hat F$ from $G_{\mu I}$ whereas the Wess-Zumino
coupling contains $B_{\mu I}$ RR gauge fields.

Fluctuations around the classical 
instanton saddle point are taken into account 
by the elliptic genus for $N$ $D1$-branes, which is
equivalent to the one of a single $D1$-brane wrapped $N$
times over $T^2$. This counting procedure is captured by
the action of the Hecke operator on the elliptic genus 
$H_N[{\cal A}](U)$ [see also eq. (8)].

As an example we write the $F_+^4$ coupling of (\ref{intFV})
as D-instanton sum:
\begin{equation}\label{ffff}
\langle F_+^4\rangle_{\rm inst}=\frac{\partial^4}{\partial s_+^4}\sum
\frac{1}{\sqrt{\det(\hat G+\hat F)}}\, e^{-S_{BI}}\, {\cal A}(\tilde{U})
\end{equation}
where $\sqrt{\det(\hat G+\hat F)}=k T_2|p-\frac{i s_+}{2\pi\sqrt{2 T_2 U_2}}|$.

In analogy to semiclassical instanton 
calculations the correlation
function in an instanton background is obtained
by saturating fermionic zero modes and integrating over the
moduli space of instantons. In this case
the instanton moduli space is provided 
by the heterotic matrix string model \cite{lowe}, describing
a worldsheet $O(N)$ two dimensional gauge theory. 
In the infrared limit this gauge theory flows
to a $(8,0)$ supersymmetric  $S_N\times {\bf Z}_2^N$ 
orbifold conformal field theory. The elliptic genus
for $S_N$ symmetric orbifolds is naturally described
by the action of the $N$'th Hecke operator on the elliptic genus
\cite{dvv}. 
\subsection{One-Loop corrections to $R^2$}
Let us consider the $CP$-even $R_{\mu\nu\rho\sigma}
R^{\mu\nu\rho\sigma}$ coupling in eight dimensions.
On the heterotic side the one-loop correction vanishes, whereas
the type I correction is non-trivial as we will show in the following.
The graviton vertex operator in the zero ghost picture is:
\begin{equation}\label{Ivert}
V_{\rm grav}=\epsilon_{\mu\nu}\ :
\Big(\bar {\partial} X^{\nu}-\frac{1}{4} \bar S^a\gamma^{\nu\rho}_{ab}
\bar S^{b} k_{\rho}\Big) 
\Big({\partial} X^{\mu}-\frac{1}{4} S^{a}\gamma^{\mu\sigma} S^b  
k_{\sigma}\Big) e^{i k X}:
\end{equation}
Although the kinematic structure of a two point graviton amplitude vanishes
due to the on shell constraints we can still calculate its 
four derivative gravitational coupling $\Delta_{\rm grav}^I$.
A three point amplitude which e.g. includes a modulus and
two gravitons will then give a non-vanishing kinematic structure multiplied
by the derivative of the same coupling with respect to the modulus 
$\partial_U \Delta_{\rm grav}^I$ \cite{ab}. 

The torus amplitude vanishes
since the fermionic zero modes cannot be saturated.
But we get non-vanishing contributions from ${\cal K}$, ${\cal A}$ and
${\cal M}$.
In order to extract the order ${\cal O}(k^4)$ term of the 
amplitude, we  have to contract the eight fermions since otherwise
we will get a zero result due to Riemann identities. 
For non-oriented surfaces there are
additional contractions between chiral and anti-chiral fermions
$\langle \psi(z)\bar \psi(\bar w)\rangle_{\sigma}=G_F(z,I_{\sigma}(w))$
with the involution
$I_{{\cal A}}(w)=I_{{\cal M}}(w)=I_{{\cal K}}-\frac{\tau}{2}
=1-\bar w$ and
$G_F(z,w)=\frac{i}{2}
\frac{\theta_{\alpha}(z-w,\tau)\theta_1'(\tau)}{
\theta_1(z-w,\tau) \theta_{\alpha}(\tau)}$.

Using the same Riemann identity as before
and taking the sum over worldsheets $\sigma$
gives for the type I one loop  correction to ${\cal R}^2$:
\begin{equation}\label{rI}
\Delta^I_{\rm grav}=
\frac{V^{(8)}T_2}{2^{8} \pi^5}
\Gamma(3) c_0\frac{U_2^3}{T_2^3}\sum_{j,p\neq 0} \frac{1}{|j-p U|^6}
\end{equation}
where the coefficient $c_0=\frac{N^2-N+2^4}{2}$ arises after taking the
sum over worldsheets $\sigma$. The coupling coincides with 
the one of $F_+^2 F_-^2$. In the decompactification limit 
$\Delta_{\rm grav}^I$ disappears,
in agreement with heterotic-type I duality in ten dimensions \cite{ts}.

Duality relates this term to a one--loop correction to ${\cal R}^2$ 
on the heterotic side. Since such a term does not exist, we conclude that
on the type I side it is a combination of ${\cal R}^2$ and 
$\Delta_{F_+^2 F_-^2}$,
which corresponds to the heterotic ${\cal R}^2$ term.
In particular this combination is such that no one--loop
correction to ${\cal R}^2$ is predicted on the heterotic side.
A similar observation was recently made with the ${\cal R}^2$ correction
in the duality of heterotic--type IIA \cite{costas}.

{\bf Acknowledgements:}
I am grateful to C. Angelantonj, C. Bachas, E. Kiritsis for 
interesting discussion and S. Stieberger for collaboration and comments.  
I would like to thank the organizers of the TMR meeting.

\end{document}